\documentclass[conference,10pt]{IEEEtran}

\usepackage[ruled,vlined]{algorithm2e}
\usepackage{bm} 

\usepackage{epsfig}
\usepackage{amssymb} 
\usepackage[tbtags]{amsmath} 
\usepackage{latexsym}
\usepackage{euscript}
\usepackage{graphics,eepic,epic,psfrag}
\usepackage{enumitem}
\usepackage{xhfill}
\usepackage{relsize}
\usepackage{exscale}
\usepackage{breqn}

\usepackage[font = footnotesize]{caption}
\usepackage{subcaption}
\usepackage{graphicx}
\usepackage{cite}  
\usepackage{xpatch}
\usepackage{algorithmicx}
\usepackage{mathtools}

\DeclarePairedDelimiter\floor{\lfloor}{\rfloor}
\usepackage{romannum}
\usepackage{comment}

\newcommand\numberthis{\addtocounter{equation}{1}\tag{\theequation}}

\ifCLASSINFOpdf
\else
\fi

\IEEEoverridecommandlockouts
\def\footnoterule{\relax%
  \kern-5pt
  \hbox to \columnwidth{\vrule width 0.5\columnwidth height 0.4pt\hfill}
  \kern4.6pt}

\begin{document}
%

\title{Active Sensing for Localization with \\ Reconfigurable Intelligent Surface}


\author{\IEEEauthorblockN{Zhongze Zhang,
Tao Jiang, and
Wei Yu}
\IEEEauthorblockA{Department of Electrical and Computer Engineering, University of Toronto \\ 
\{ufo.zhang,taoca.jiang\}@mail.utoronto.ca,  weiyu@ece.utoronto.ca}
}

\maketitle

%




\begin{abstract}


This paper addresses an uplink localization problem in which the base station (BS) aims to 
locate a remote user with the aid of reconfigurable intelligent surface (RIS). 
This paper proposes a strategy in which the user transmits pilots over
multiple time frames, and the BS adaptively adjusts the RIS reflection coefficients 
based on the observations already received so far in order to produce an accurate 
estimate of the user location at the end. This is a challenging active sensing problem for which finding an optimal
solution involves a search through a complicated functional space whose 
dimension increases with the number of measurements. In this paper, we show that the long short-term memory (LSTM) network can be used to exploit the latent temporal correlation between measurements to
automatically construct scalable information vectors (called hidden state) 
based on the measurements.  Subsequently, the state vector can be mapped to the
RIS configuration for the next time frame in a codebook-free fashion via a deep
neural network (DNN). After all the measurements have been received, a final DNN
can be used to map the LSTM cell state to the estimated user equipment (UE)
position. Numerical result shows that the proposed active RIS design results
in lower localization error as compared to existing active and nonactive methods. The proposed
solution produces interpretable results and is generalizable to early stopping in the sequence of sensing stages.
\end{abstract}




%

\section{Introduction}

Reconfigurable intelligent surface (RIS) is a planar surface consisting of a large number of passive elements, with each element capable of altering the phase of the incident electromagnetic wave with very low power consumption\cite{basar2019wireless}.
The technology is envisioned as one of the enabling technologies for 6G localization due to its ability to:
\romannum{1})
establish a reliable reflected link when the direct path between the transceivers is weak or blocked, and \romannum{2})
provide new anchor points without needing expensive RF chains  \cite{6gwhitepaper,panpan}. 
In the area of RIS-aided localization, fruitful research progress has been made in
the direction of theoretical bounds and algorithms \cite{siso, nearfield,Ubiquitous,Ubiquitous2,wang2021joint, nguyen,hejiguang,liu2020reconfigurable}.


This paper considers an uplink localization problem in the presence of RIS,
where a single user repeatedly transmits pilot symbols and the base station
(BS) receives the pilots through the reflection at the RIS and determines the location
of the user based on the received pilots. 
Specifically, this paper investigates a scenario in which the BS can control 
the reflection coefficients at the RIS and tackles the problem of designing the RIS 
configuration in an \emph{active} manner to
minimize localization error. By active sensing, we mean that the RIS
reflection coefficients are sequentially designed as a function of previous
measurements. As a result, the RIS can be used to focus the beam progressively to
locate the user over time as more measurements become available. This is however
a challenging problem as a solution needs to be able to optimize over the
complicated functional landscape from the sequence of observations to the RIS reflection coefficients, while being scalable in the number of measurements. 



In the context of localization algorithms in RIS-assisted network, the design of RIS configuration to enhance localization accuracy is of great interest. 
The authors in \cite{nearfield} devise three classes of RIS profiles and evaluated the impact of each class on the 3D position error bound of the user equipment (UE).
Two of the three classes are heuristically designed to produce narrow/broad beams, 
and the remaining class consists of random RIS profiles. 
In \cite{Ubiquitous,Ubiquitous2}, the indoor localization accuracy is improved by designing a set of RIS profiles that enlarges the differences between the received signal strength (RSS) values of adjacent locations. The set of RIS profiles is designed via local and global search methods.
The authors in 
\cite{wang2021joint} consider joint localization and communication optimization problem and design the appropriate RIS profiles to improve localization accuracy and transmission throughput. 
In \cite{nguyen}, the authors consider a fingerprinting localization problem enabled by an RIS. In particular, the authors employ machine learning to identify a subset of RIS profiles from a codebook to generate a more diverse set of fingerprints to improve fingerprint matching accuracy. 
All of the above works have demonstrated the importance of designing/selecting an appropriate set of RIS configurations and the associated performance gain. However, \emph{active} design of RIS configuration, which could further improve the performance, has not been considered.



In this paper, we focus on the active design of RIS reflection coefficients 
based on the sequential historical measurements of the environment 
to improve the localization accuracy. 
Designing such an adaptive sequence of RIS patterns is a challenging problem. To make the problem tractable, the authors in \cite{hejiguang} design a hierarchical codebook for the RIS that enables adaptive bisection search over the angular space in a 2D localization setting. 
However, a hierarchical codebook-based method is not necessarily the best approach. For example, in the context of angle of arrival (AoA) estimation problem, the performance of the hierarchical codebook-based method is severely restricted by the quality of the codebook and is far from optimal \cite{aoa_adaptive}. 
In another work on active sensing with RIS \cite{liu2020reconfigurable}, 
the authors use gradient descend method to optimize RIS configuration
with a goal of minimizing the Cram\'er-Rao lower bound (CRLB) in each step. 
However, the method is based on first estimating the fading coefficients and AoAs instead of the location, which results in significant pilot overhead.

This paper proposes a learning-based localization solution, where a sequence of RIS configurations is adaptively designed based on the historical measurements in a codebook-free fashion. In particular, 
we use 
the long short-term memory (LSTM) based network, for its ability to capture the temporal relationship between different measurements over a long period\cite{LSTM}.
We use a chain of LSTM cells, corresponding to the sequence of measurements.
At each measurement time frame, an LSTM cell accepts new measurement of the environments, i.e., received pilots, and automatically uses the newly
received measurement (along with historical measurements) to update a hidden state vector of fixed dimension. 
Subsequently, a deep neural network (DNN) is used to extract a correct representation of the knowledge of location so far from the hidden state
to design the RIS profile for the next time frame. 
After multiple measurements, the cell state of the final LSTM cell is passed through another DNN to obtain
the estimated UE position.
Numerical result shows that the proposed active RIS profile design achieves lower localization error as compared to the existing nonactive and active RIS profile design.








\section{System Model and Problem Formulation}
\label{sec.background}

\subsection{System Model}

\begin{figure}[t]
\centering
\includegraphics[width=\columnwidth]{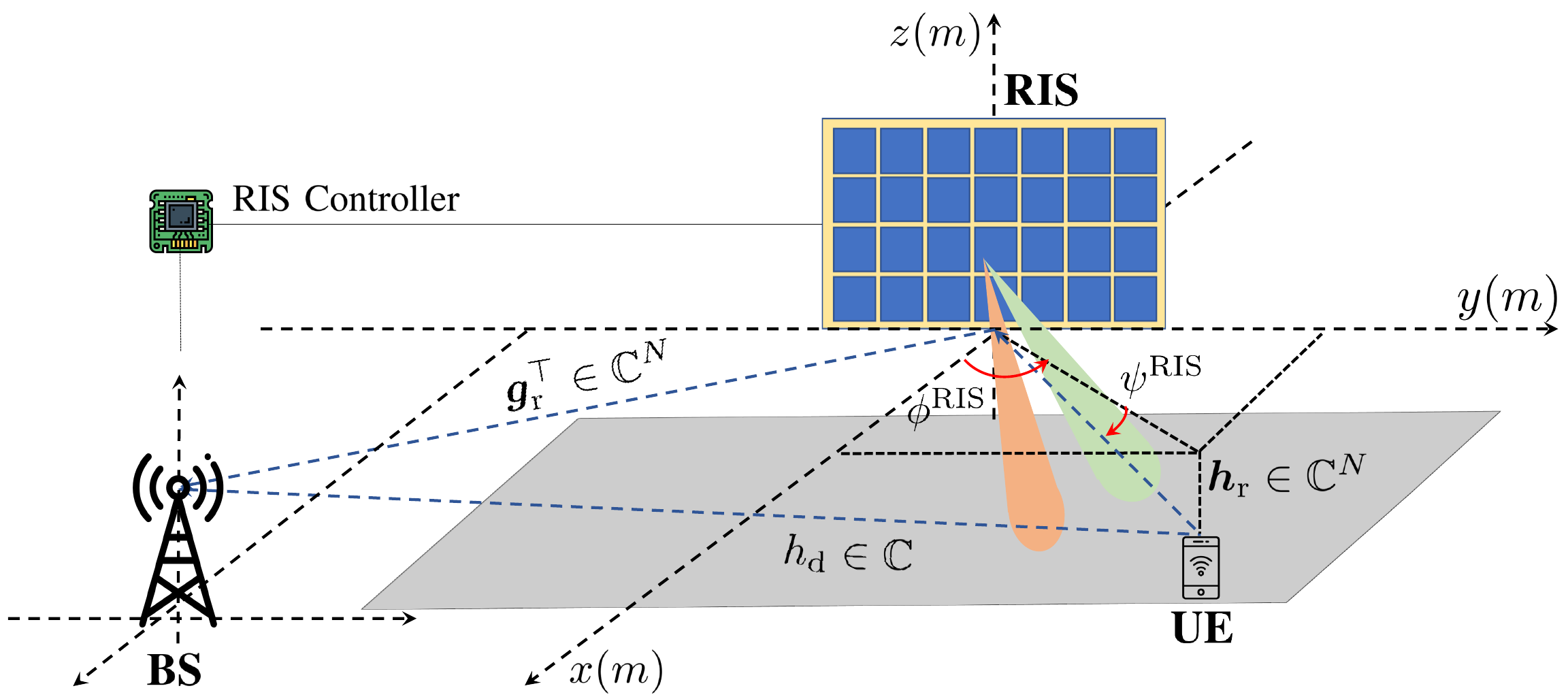}%
\caption{RIS-assisted network.}%
\label{fig.setup_intro}%
\vspace{-0.4em}
\end{figure}

We consider a localization problem in an RIS-assisted system with a single-antenna BS, a single-antenna UE, and a planar RIS. The BS and RIS are placed as in Fig.~\ref{fig.setup_intro} to localize the potential users in the area. 
We assume that the position of the BS and the RIS are known. By adopting the Cartesian coordinate system, let $\bm{p}^{\rm BS} = [x^{\rm BS},y^{\rm BS},z^{\rm BS}]^\top$ and $\bm{p}^{\rm RIS} = [x^{\rm RIS},y^{\rm RIS},z^{\rm RIS}]^\top$ denote the position of the BS and the RIS respectively.
The unknown UE position is denoted as $\bm{p} = [x,y,z]^\top$.

The reflection coefficients of the RIS are controlled by an RIS controller which receives controlling signals from the BS. Let $N$ be the number of reflection coefficients at the RIS. Then the RIS reflection coefficient is denoted as
\begin{equation}
    \bm{\theta} = [e^{j\delta_1},  e^{j\delta_2}, \cdots , e^{j\delta_{N}}]^\top\in\mathbb{C}^{N},
\end{equation}
with $\delta_n \in [0,2\pi)$ as the phase shift of the $n$-th element.

We adopt a \emph{block-fading} model in which the channels are assumed to be constant across multiple time frames within a coherence period, then change independently in subsequent coherence periods. 
As shown in Fig.~\ref{fig.setup_intro}, $h_{\rm{d}}\in\mathbb{C}$ denotes the direct channel from the BS to the UE, $\bm{h}_{{\rm{r}}}\in\mathbb{C}^{N}$ denotes the reflection channel from the RIS to the UE, and $\bm{g}_{\rm{r}}^\top \in\mathbb{C}^{N}$ denotes the channel from the BS to the RIS. We further assume that the reflection channel $\bm{h}_{{\rm r}},\bm{g}_{\rm r}$ and the direct channel $h_{\rm d}$ follow Rician fading model:
\begin{subequations} 
\begin{align}
\bm{h}_{{\rm r}} =\; & \kappa\left(\sqrt{{\epsilon}/({1+\epsilon})} \bm{\tilde{h}}_{{\rm r}}^{\rm LOS} + \sqrt{{1}/({1+\epsilon})}\bm{\tilde{h}}_{{\rm r}}^{\rm NLOS}\right),\\
\bm{g}_{\rm r} =\; & \xi\left(\sqrt{{\epsilon}/({1+\epsilon})}\bm{\tilde{g}}_{\rm r}^{\rm LOS} + \sqrt{{1}/({1+\epsilon})}\bm{\tilde{g}}_{\rm r}^{\rm NLOS}\right),\\
h_{\rm d}  =\;& \rho\left(\sqrt{{\epsilon}/({1+\epsilon})} \tilde{h}_{{\rm d}}^{\rm LOS} + \sqrt{{1}/({1+\epsilon})}\tilde{h}_{{\rm d}}^{\rm NLOS}\right),
\end{align}
\end{subequations}
where $\rho$ denotes the pathloss between the BS and the UE, and $\kappa$ and $\xi$ denote the path losses between the RIS and the UE/BS. 
Here, $\tilde{h}^{\textrm{NLOS}}_{{\rm d}}$,
$\bm{\tilde{h}^{\textrm{NLOS}}}_{{\rm r}}$ and $\bm{\tilde{g}}_{\rm r}^{\textrm{NLOS}}$ denote the non-line-of-sight components and their entries are generated independently according to $\mathcal{C}\mathcal{N}(0,1)$. 
The line-of-sight component of the reflection channel contains angular information about the location of UE. 
For example, $\bm{\tilde{h}}_{{\rm r}}^{\rm LOS}$ is a function of UE location and RIS location. Let $\phi^{\rm RIS}$ and $\psi^{\rm RIS}$ denote the azimuth and elevation angle of arrival from the UE to the RIS. The line-of-sight component of $\bm{{h}}_{{\rm r}}$ can be written as
\begin{equation}
    \bm{\tilde h}_{{\rm r}}^{\rm LOS} = \bm{a}^{\rm RIS}(\phi^{\rm RIS}, \psi^{\rm RIS}),
\end{equation}
\noindent where the steering vector of the $n$-th element of the RIS can be expressed as \cite{taojournal}
\begin{align*}
[ \bm{a}^{\rm RIS} & (\phi^{\rm RIS}, \psi^{\rm RIS})]_n \\ & =
e^{\scriptsize \begin{array}{ll}j \dfrac{2\pi d_{\rm R}}{\lambda_c}  \end{array} \{ v_1(n, N_c) {\rm sin}(\phi^{\rm RIS}){\rm cos}(\psi^{\rm RIS}) + v_2(n, N_c) {\rm sin}(\psi^{\rm RIS}) \}},\numberthis \label{eqn}
\end{align*}
where $d_{\rm R}$ is the distance between two reflective elements of the RIS, $\lambda_c$ is the carrier wavelength, and $v_1(n,N_c) = {\rm mod}(n-1,N_c)$ and $v_2(n, N_c) = \floor*{\frac{n-1}{N_c}}$. Here, $N_c$ is the number of columns of the RIS. 

Recall that $\bm{p}^{\rm RIS} = [x^{\rm RIS},y^{\rm RIS},z^{\rm RIS}]^\top$ denote the position of the RIS and $\bm{p} = [x,y,z]^\top$ denote the position of the UE, we have the following\cite{taojournal}:
\begin{subequations} 
\begin{align}
{\rm sin}(\phi^{\rm RIS}){\rm cos}(\psi^{\rm RIS}) & =({y-y^{\rm RIS}})/{d},\\
{\rm sin}(\psi^{\rm RIS})  &=({z-z^{\rm RIS}})/{d},
\end{align}
\end{subequations}
where $d$ is the distance between the RIS and the UE.

Let $\eta^{\rm RIS}$ and $\vartheta^{\rm RIS}$ denote the azimuth and elevation angle of departures (AoD) from the RIS to the BS. 
Since the BS only has a single antenna, the line-of-sight component of $\bm{g}_{\rm r}$ is given by 
\begin{equation}
    \bm{\tilde g}_{\rm r}^{\rm LOS} = ~ \bm{a}^{\rm RIS}(\eta^{\rm RIS}, \vartheta^{\rm RIS})^{\rm H},
\end{equation}
where the line-of-sight component of $h_{\rm d}$ is given by
\begin{equation}
    \tilde h_{\rm d}^{\rm LOS} = 1.
\end{equation}

\subsection{Signal Transmission}

When there is a localization request, the UE 
sends a sequence of $T$ uplink pilot symbols to the BS over $T$ time frames. 
Let $x_t\in\mathbb{C}$ be the pilot symbol to be transmitted from the UE to the BS at the $t$-th time frame. The BS receives a combination of the signal from the direct path and the signal reflected off the RIS, so the received pilots at the BS can be expressed as
\begin{equation}
    y_t(\bm{\theta}_{t}) = \sqrt{P_u}(h_{\rm d} + \bm{v}_{\rm r}^\top \bm{\theta}_{t})x_t + n_t ,~~t = 0,\cdots, T-1,
\end{equation}
where $P_u$ is the uplink transmission power, $\bm{v}_{\rm r} = \text{diag}(\bm{h}_{{\rm{r}}})\bm{g}_{\rm{r}}^\top\in\mathbb{C}^{N}$ is the cascade channel between the BS and the UE through the reflection at the RIS, and $n_t \sim\mathcal{C}\mathcal{N}(0,\sigma_u^2)$ is the uplink additive white Gaussian noise. 
Here, the received pilot is a function of the RIS configuration at the $t$-th time frame $\bm{\theta}_{t}$,  where the reflection coefficients can be configured randomly or designed according to some criteria to enable a better localization outcome.


\subsection{Problem Formulation}

The goal of the localization problem is to estimate the unknown UE position $\bm{p}$ based on the $T$ observations $[{y}^{(t)}]_{t=0}^{T-1}$, and the known BS and RIS positions. 
The design of RIS configuration is of critical interest here. It has been shown in \cite{nguyen,Ubiquitous,Ubiquitous2} that by strategically selecting a fixed set of RIS configurations, a more favourable RSS distribution can be obtained, which allows improvement in localization accuracy. 
However, most existing works are based on the \textit{fixed} sensing framework, which passively collects all the observations of the environment according to a fixed set of RIS configurations.
In essence, the RIS probes the search area using fixed beams along multiple random directions.
 Here, we instead propose an \textit{active} sensing framework to gradually narrow down the searching area using more directional beams. 



Specifically, we consider the following active localization setup. 
In the $t$-th time frame, the BS designs the next RIS configuration $\bm{\theta}_{t+1}$ based on the existing observations, which is used to make the next measurement ${y}_{t+1}$ in the $(t+1)$-th time frame. Thus, the design of RIS configuration is a function of historical measurements:
\begin{equation}\label{function_g}
    \bm{\theta}_{ t+1} = \mathcal{G}_t(\{{y}_\tau\}_{\tau=0}^{t}), ~ t = 0,\cdots, T-1.
\end{equation}
As no prior observation exists when $t < 0$, the first set of RIS configurations $\bm{\theta}_{0}$ is produced via function $\mathcal{G}_{-1}(\emptyset)$. The function accepts an empty set as input and always produces the same initialization of RIS configuration. 

The estimated UE position $\hat{\bm{p}}$ can be written as a function of all $T$ historical observations. 
\begin{equation} \label{function_f}
    \hat{\bm{p}}  = \mathcal{F}(\{{y}_t\}_{t=0}^{T-1}).
\end{equation}
The location estimation problem can now be formulated as:
\begin{subequations} \label{prob_formulation}
\begin{align}
\underset{\scriptsize \begin{array}{ll} \{\mathcal{G}_t(\cdot)\}_{t=0}^{T-1}, \mathcal{F}(\cdot)  \end{array} }{\textrm{min}} & \mathbb{E}\left[ \| \hat{\bm{p}}- \bm{p} \|_2^2 \right]\\
\textrm{subject to} \quad & \left|[\bm{\theta}_{t}]_n\right|= 1, ~\forall n,t, \\
& \bm{\theta}_{t+1} = \mathcal{G}_t(\{{y}_\tau\}_{\tau=0}^{t}), ~t = 0, \cdots, T-1 ,\\
& \hat{\bm{p}}  = \mathcal{F}(\{{y}_t\}_{t=0}^{T-1}).
\end{align}
\end{subequations}

Solving the optimization problem (\ref{prob_formulation}) is challenging, as
the problem amounts to optimizing the function expression
$\{\mathcal{G}_t(\cdot)\}_{t=0}^{T-1}$ in (\ref{function_g}) and $\mathcal{F}(\cdot)$ in
(\ref{function_f}). To make the problem more tractable, a common approach is to
select the RIS configuration adaptively from a predefined set of RIS patterns,
namely the \emph{codebook}, based on heuristics. For example, \cite{hejiguang}
considers a 2D RIS-assisted localization problem, where the sequence of RIS
patterns is heuristically selected from hierarchical codebooks, based on
feedback from the UE. However, a codebook-based approach is not ideal as the
freedom of designing RIS configuration is restricted. 

In this work, we propose to employ a neural network to parameterize the function $\mathcal{F}(\cdot)$ and $\mathcal{G}_t(\cdot)$ to adaptively design a sequence of RIS configurations, 
without restricting the design space of RIS coefficients to a pre-defined codebook. 




\section{Proposed Active Localization Framework}
\label{sec.rnn}

In this section, we introduce the deep learning approach to solving the active localization problem discussed in (\ref{prob_formulation}). 
Here in problem (\ref{prob_formulation}), the next RIS configuration $\bm{\theta}_{t+1}$ is designed based on historical measurements up to $t$, i.e., 
$ \{{y}_\tau\}_{\tau=0}^{t}$. 
The main design challenge is that as the dimension of historical observations increases linearly with $t$, it is difficult to use the entire history of measurements to design the next RIS configuration for large $t$. 
Thus, the key design question for the neural network is how to extract
useful information from the historical measurements and how to map it to 
an information vector of fixed dimension.

In this paper, we develop an LSTM network capable of automatically constructing information vector of fixed dimension from existing measurements and extracting temporal features and long-term dependencies from a sequence of temporal input. 
The specific LSTM network developed here is similar to the one in
\cite{sohrabi2021active} in which the information vector is the hidden state,
but with modifications to the neural network architecture tailored to the localization problem in an RIS-assisted network. 
We also propose a new loss function to make
the LSTM network generalizable to the number of time frames.
\begin{figure}[!t]
\centering
  \includegraphics[width=0.5\textwidth]{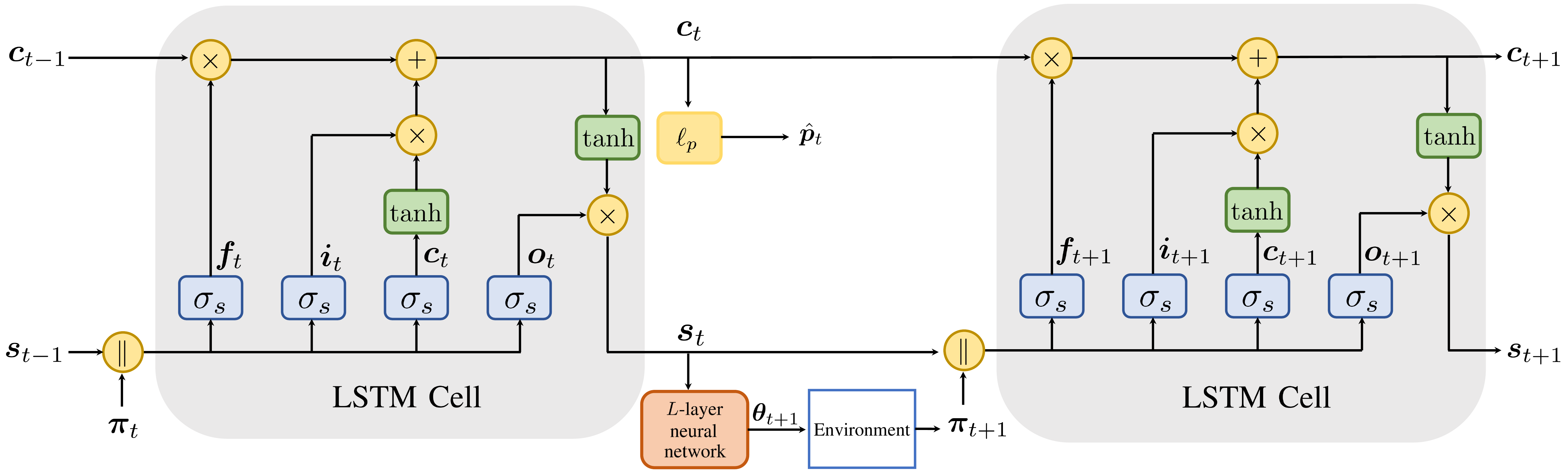}%
  \caption{Proposed active localization framework.}
  \label{fig.lstmcell}%
\end{figure}
\subsection{Neural Network Architecture}
The overall neural network architecture is shown in Fig.~\ref{fig.lstmcell}. 
At the $t$-th time frame, an LSTM cell takes new feature as input to update the hidden state vector $\bm{s}_t$ and the cell state vector $\bm{c}_t$. The new feature, denoted as $\bm{\pi}_t$, can be the RSS $|{y}_t|^2$, or the real and imaginary component of received pilots $[\mathcal{R}({y}_t), \mathcal{I}({y}_t)]$, depending on the constraints of the detecting hardware at the receiver.
We use $\circ$ to denote the element-wise product.
The updating rules of the state vectors are as follows:
\begin{subequations} 
\begin{align}
\bm{c}_t = \; & \bm{f}_t \circ \bm{c}_{t-1} + \bm{i}_t \circ {\rm tanh}\left(\bm{u}_c\bm{\pi}_t + \bm{w}_c\bm{s}_{t-1}\right),\\
\bm{s}_t = \; & \bm{o}_t \circ {\rm tanh}(\bm{c}_t),
\end{align}
\end{subequations}
where $\bm{u}_c$ and $\bm{w}_c$ are linear layers with a number of fully connected units. Here, $\bm{f}_t$, $\bm{i}_t$, and $\bm{o}_t$ are the activation vectors of the forget gate, input gate and output gate within the LSTM cell respectively. The element-wise updating rules of the different gates are as follows:
\begin{subequations}
\begin{align}
\bm{f}_t =\;& \sigma_s\left(\bm{u}_f\bm{\pi}_t+\bm{w}_f\bm{s}_{t-1}\right),\\
\bm{i}_t =\;& \sigma_s\left(\bm{u}_i\bm{\pi}_t+\bm{w}_i\bm{s}_{t-1}\right),\\
\bm{o}_t =\;& \sigma_s\left(\bm{u}_o\bm{\pi}_t+\bm{w}_o\bm{s}_{t-1}\right),
\end{align}
\end{subequations}
\noindent where $\sigma_s(x) = 1/(1+e^{-x})$ is the sigmoid function, and $\bm{u}_f$, $\bm{u}_i$, $\bm{u}_o$, $\bm{w}_f$, $\bm{w}_s$ and $\bm{w}_o$ are linear layers. As per convention, the initial value of the cell state $\bm{c}_0$ is obtained by setting $\bm{s}_{-1} = \bm{c}_{-1}=\bm{0}$, and ${y}_{0} = 1$.


We design the RIS configuration for time frame $t+1$ based on the hidden state vector $\bm{s}_t$. The hidden state vector is used as the input to a fully connected neural network of $L$ layers to design the RIS configuration for the next time frame, 
\begin{equation}
    \bm{\bar{\theta}}_{t+1} = \beta_L(\bm{A}_L \beta_{L-1}(\cdots \beta_1(\bm{A}_1\bm{s}_t+\bm{b}_1)\cdots)+\bm{b}_L),
\end{equation}
\noindent where $\beta_l$, $l\in \{1,\cdots,L\}$ is the activation function of the $l$-th layer, i.e., ${\rm relu}(x) = {\rm max}(0,x)$, $\{\bm{A}_l\}_{l=1}^{L}$ and $\{\bm{b}_l\}_{l=1}^{L}$ are sets of trainable weights and biases. 
Here, $\bm{\bar{\theta}}_{t+1}$ contains the real and imaginary components of the RIS reflection coefficients
\begin{equation}
    \bm{\bar{\theta}}_{t+1} = [ \mathcal{R}(\bm{{\theta}}_{t+1})^\top, \mathcal{I}(\bm{{\theta}}_{t+1})^\top]^\top.
\end{equation}
The dimensions of $\bm{A}_L$ and $\bm{b}_L$ are designed to ensure the output is of correct dimension, such that $\bm{\bar{\theta}}_{t+1} \in\mathbb{R}^{2N}$.
To enforce unit modulus constraint on each element of the RIS, an element-wise normalization is performed
\begin{align*}
 \hspace*{-0.1em}[\bm{\theta}_{t+1}]_n= \: & {[\mathcal{R}(\bm{{\theta}}_{t+1})]_{n}}/{\sqrt{[\mathcal{R}(\bm{{\theta}}_{t+1})]_{n}^2+[\mathcal{I}(\bm{{\theta}}_{t+1})]_{n}^2}} \\
& 
+j{[\mathcal{I}(\bm{{\theta}}_{t+1})]_{n}}/{\sqrt{[\mathcal{R}(\bm{{\theta}}_{t+1})]_{n}^2+[\mathcal{I}(\bm{{\theta}}_{t+1})]_{n}^2}}. \numberthis \label{eqn}
\end{align*}






While the hidden state vector $\bm{s}_t$ is used to design the RIS configuration for the next time frame, the cell state vector $\bm{c}_T$ is used to obtain the estimated UE position at the $T$-th time frame.
After $T$ time frames, the final estimated UE position $\hat{\bm{p}}_T$ is obtained through a fully connected neural network, 
based on the final cell state $\bm{c}_T$:
\begin{equation}
    \hat{\bm{p}}_T = \bm{\ell}_p\bm{c}_T,
\end{equation}
where $\bm{\ell}_p$ denotes a neural network with linear layers. 
Here, we find that a linear network already works well as the final DNN.

\subsection{Loss Functions}

To train the LSTM network, we employ Adam optimizer \cite{adam} to minimize the average mean squared error between the estimated position $\hat{\bm{p}}_T$ and the true position as follows
\begin{equation}\label{pt_mse}
    \begin{split}
    \mathbb{E}\left[ \| \hat{\bm{p}}_T- \bm{p} \|_2^2 \right].
    \end{split}
\end{equation}
\noindent Here, the choice of loss function in (\ref{pt_mse}) encourages the LSTM network to design a series of $T$ RIS configurations to minimize localization error at the final time frame. 
We note that this loss function only accounts for the estimation error at the final stage. This is a good choice since it gives the neural network freedom to design the sensing strategy across the entire $T$ measurement stages.


In some cases, we may need to have earlier stopping in the sequence of sensing stages, e.g., due to some latency constraints. This requires the neural network to output the best estimates of the location before the final stage, but this is not considered in the loss function \eqref{pt_mse}. For this scenario, we propose an alternative loss function as follows:
\begin{equation}\label{avgpt_mse}
    \begin{split}
    \mathbb{E}\left[ \dfrac{1}{T} \sum_{t = 1}^{T} \| \hat{\bm{p}}_t- \bm{p} \|_2^2 \right],
    \end{split}
\end{equation}
where $\hat{\bm{p}}_t = \bm{\ell}_p\bm{c}_t$. By minimizing the average localization error across $t$, we can encourage the LSTM network to design RIS reflection coefficients to reduce estimation error at each time frame. The new loss function in (\ref{avgpt_mse}) helps the learning framework to achieve good generalizability with respect to the number of stages.


\section{Numerical results}
\label{sec.numerical_irs}
\subsection{Simulation Environment}
In simulations,
the BS is located at $\bm{p}^{\rm BS} = (40m,-40m,-10m)$ and an $8\times8$ RIS is located at $\bm{p}^{\rm RIS} = (0m,0m,0m)$. 
The unknown user locations $\bm{p}$ are uniformly generated within a rectangular area on the $x$-$y$ plane $(20\pm15m, 0\pm35m, -20m)$. In subsequent simulation, the Rician factor $\epsilon$ is set to $10$, and $ {2\pi d_{R}}/{\lambda_c} = 1$ without loss of generality. The path-loss models of the direct and reflected paths are $32.6+36.7\log(d_1)$ and $30+22\log(d_2)$, respectively, where $d_1$ and $d_2$ denote the corresponding link distance.

\begin{table}[t]
\begin{center}
\captionof{table}{Parameters of the LSTM Network.\label{Tab:parameter}}
\begin{tabular}{| c|c |} 
\hline
\textbf{Label}& \textbf{Dimension}\\
\hline
\shortstack{\:\\$\bm{u}_c,\bm{u}_f,\bm{u}_i,\bm{u}_o$ \\ $\bm{w}_c,\bm{w}_f,\bm{w}_i,\bm{w}_o$ } &\shortstack{ $512$ \\ \quad \quad }\\
\hline
$\bm{A}_{1} \quad \; $\vline $ \; \{\bm{A}_l\}_{l=2}^{L-1}  \; $\vline $ \;\;\;\; \bm{A}_{L}$   & $512 \times 1024  \; $\vline $ \; 1024 \times 1024 \; $\vline $ \;   1024 \times 2 N$ \\
\hline
$\{\bm{b}_l\}_{l=1}^{L-1} \quad $\vline $ \quad\quad\quad \bm{b}_{L}$   & $1024 \quad\quad\quad\;\; \; $\vline $ \; \;\;\quad\quad\quad 2 N$ \\
\hline
$\bm{\ell}_{p}$   & $ 512 \times 3$\\
\hline
\end{tabular}
\end{center}
\end{table}

\subsection{Baseline Schemes}
The proposed LSTM network for adaptive RIS-assisted localization is
implemented using parameters in Table \ref{Tab:parameter}. We set  $L = 4$ and develop the model using Tensorflow \cite{tensorflow}. The model samples a total of 2,048,000 training data in 2000 epochs. 
We compare the localization performance of the proposed algorithm against the following baseline schemes. 

\emph{Fingerprint based localization with random RIS configurations}\cite{nguyen}: In this scheme, the sequence of $T$ uplink RIS configurations is non-adaptive and random. Every $1m\times 1 m$ block within the $30m \times 70m$ area is associated with a vector of RSS as its fingerprint, i.e., $\left[|{y}_1|^2,\cdots,|{y}_T|^2\right]$.
The fingerprints are stored in an offline database. A weighted $k$-nearest neighbour (wKNN) classifier is used to match online and offline measurements. Here, we set $k = 5$.

\emph{DNN with random or learned RIS configurations}:
    The sequence of $T$ uplink RIS configurations is non-adaptive. Here, we can have two different possibilities: \romannum{1}) the RIS configurations are randomly chosen, or \romannum{2}) the sequence of RIS configurations is learned from training data, but are not adaptive as a function of previous measurements. A fully connected neural network is used to map the received pilot symbols over $T$ time frames $\{ \mathcal{R}({y}_t),\mathcal{I}({y}_t)\}_{t=0}^{T-1}$ or RSS values over $T$ time frames to an estimated UE position. The dimension of the neural network is $[ 200,200,200,3 ]$.

\emph{Optimizing CRLB using gradient descent (GD)}\cite{liu2020reconfigurable}:
Here, we test the idea of designing an active sensing strategy based on minimizing
CRLB in every time step. This is an adaptive scheme based on estimation theory
rather than machine learning. 
In each time frame, 
based on received measurements, we use maximum a posterior (MAP) estimator to estimate the UE location, which is used to update the CRLB of the location error. Here, the CRLB is calculated by replacing the unknown true location of UE with the estimated one. 
Subsequently, the RIS reflection coefficients are designed using the GD algorithm to minimize the approximated CRLB for the next pilot transmission. 

\begin{figure}
\includegraphics[width=\columnwidth]{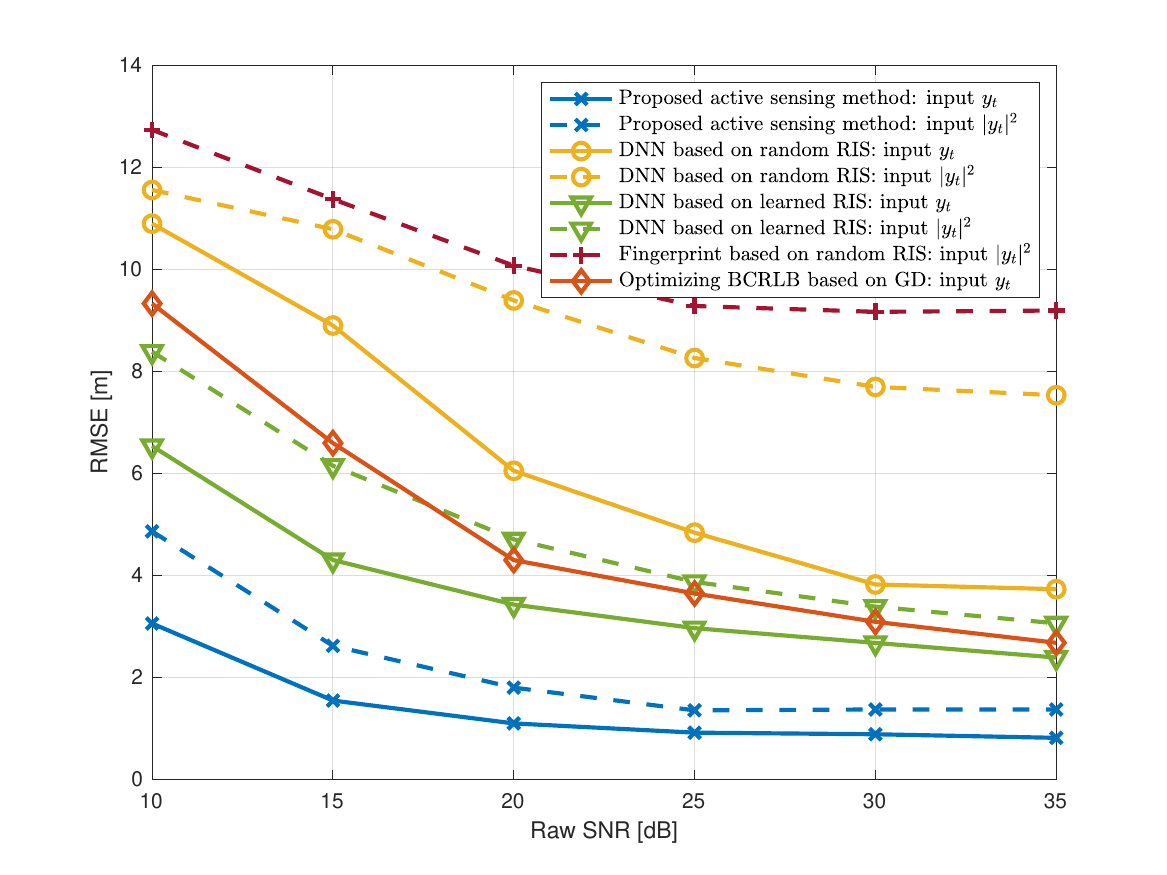}%
\caption{Localization accuracy vs. raw SNR, $N=64$, $T=6$.}%
\label{fig.loc_ris_mse_vs_snr}%
\end{figure}

\begin{figure}
\includegraphics[width=\columnwidth]{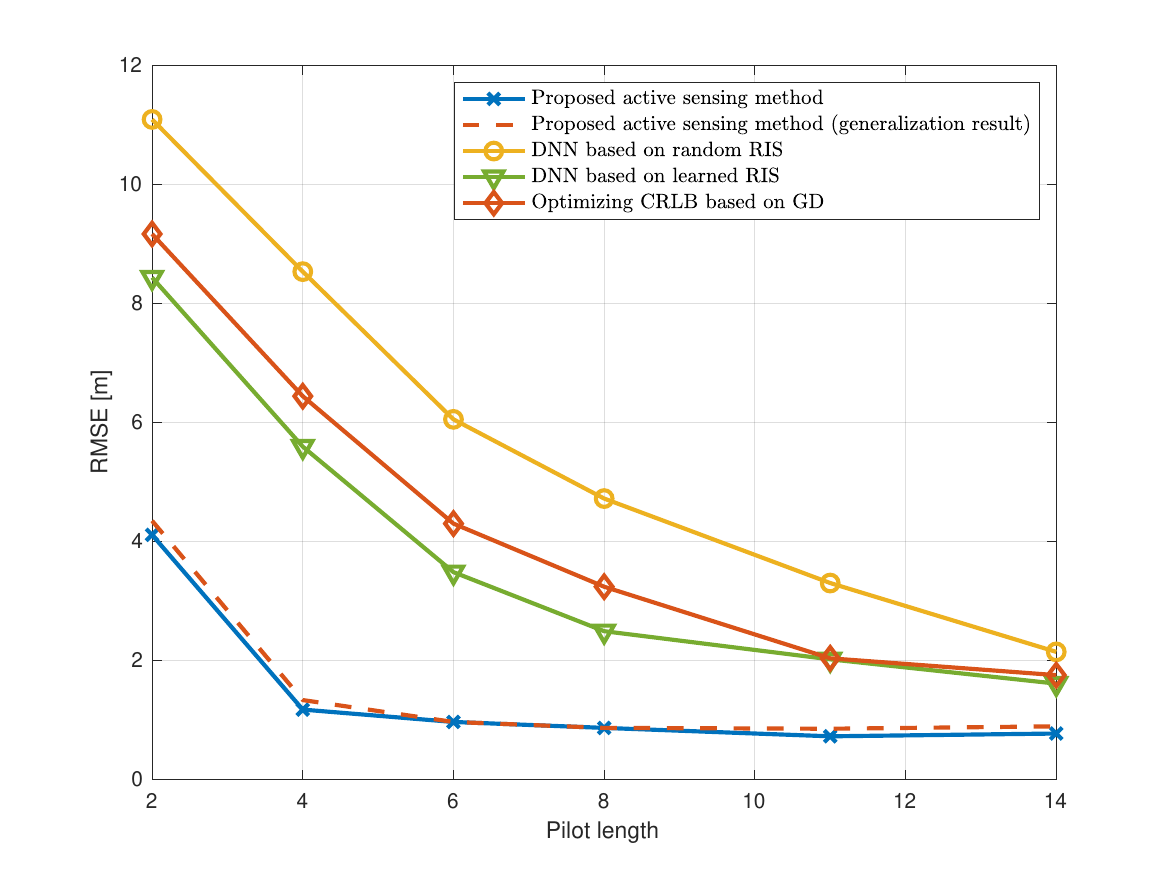}%
\caption{Localization accuracy vs. pilot length, $N=64$, SNR$=20$dB.}%
\label{fig.loc_ris_mse_vs_t}%
\end{figure}

\subsection{Simulation Results}

We first examine the localization performance versus raw SNR, i.e., $P_u = 10^{{\rm SNR}/10}$, when $T = 6$.  
In Fig.~\ref{fig.loc_ris_mse_vs_snr}, whether the input is the pilot symbols or RSS value, 
the proposed adaptive RIS design is seen to have the highest localization accuracy across different SNRs as compared to other benchmarks with non-adaptive RIS design. 
This implies that the proposed algorithm is effectively utilizing the current and historical measurements to design a more suitable RIS configuration for future time frames to minimize localization error.

Here, we point out that CRLB-minimization based adaptive RIS design does not serve as an optimal RIS design in reducing the location MSE
for the following reasons: 
\romannum{1}) an active sensing solution to (\ref{prob_formulation}) should design a series of sensing configuration jointly across $t$ to minimize the location MSE, i.e., $\{{\mathcal{G}}_{t}(\cdot)\}_{t=0}^{T-1}$ as in (\ref{prob_formulation}). However, 
the CRLB-based approach greedily treats the active sensing problem by designing individual ${\mathcal{G}}_{t}(\cdot)$ at each time frame, 
\romannum{2}) the CRLB can be a loose lower bound of the MSE, especially when SNR is low and/or the number of observations is limited\cite{crlbbound,crlbbound2} 
\romannum{3}) using the estimated UE location in computing the CRLB can introduce errors,
\romannum{4}) 
finally, the optimization of CRLB is a nonconvex optimization problem; it is difficult to find its true optimal solution.




We point out that the fingerprinting based approach along with wKNN classifier is seen to experience poor performance in this simulation setup. This is due to the randomness in the non-line-of sight Rician channel model which adversely influences the fingerprint matching accuracy. 

We next examine the localization performance with varying numbers of time frames. From Fig.~\ref{fig.loc_ris_mse_vs_t}, we observe that the proposed algorithm demonstrates a robust performance over other benchmarks. Moreover, the LSTM based neural network with the proposed training algorithm generalizes well to early stopping in the sequence of time frames. The model is trained to adaptively design RIS configurations to minimize loss function (\ref{avgpt_mse}) when $T=14$. We test the trained model when the number of time frames is fewer than $T$ without retraining, and find the localization accuracy to be as good as if the LSTM is trained for each specific $T$. 

\begin{figure}[t]
  \centering
  \begin{subfigure}{.5\columnwidth}
    \centering
    \includegraphics[width=\linewidth]{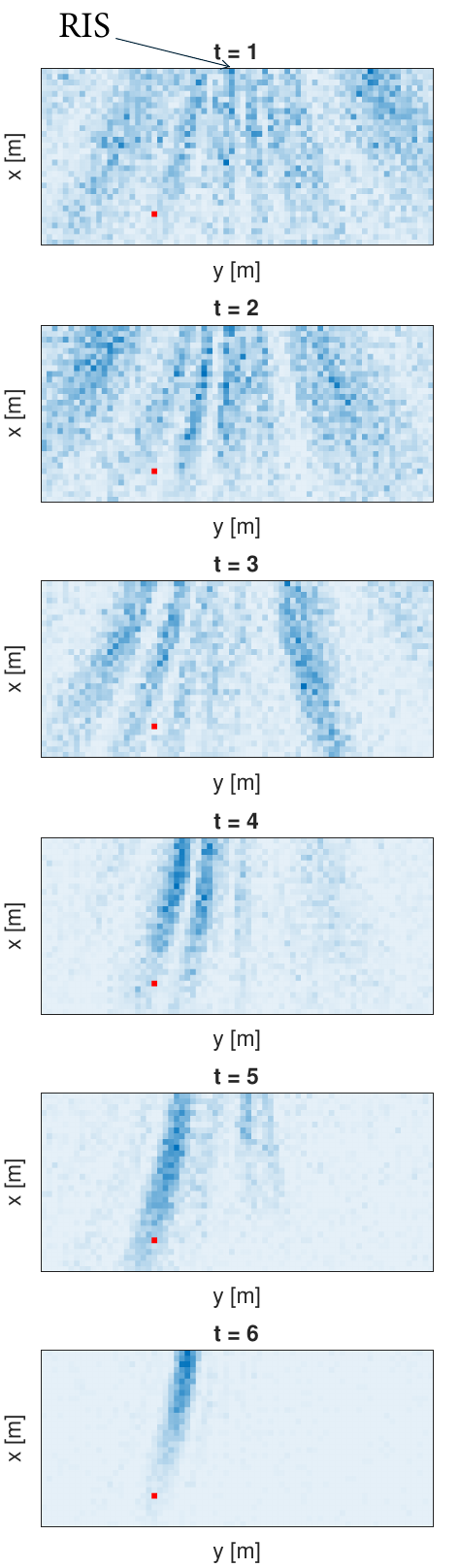}
\caption{The proposed active sensing method.}
    \label{loc_ris_interp_6_rnn}
  \end{subfigure}%
  \hfill
  \begin{subfigure}{.5\columnwidth}
    \centering
    \includegraphics[width=\linewidth]{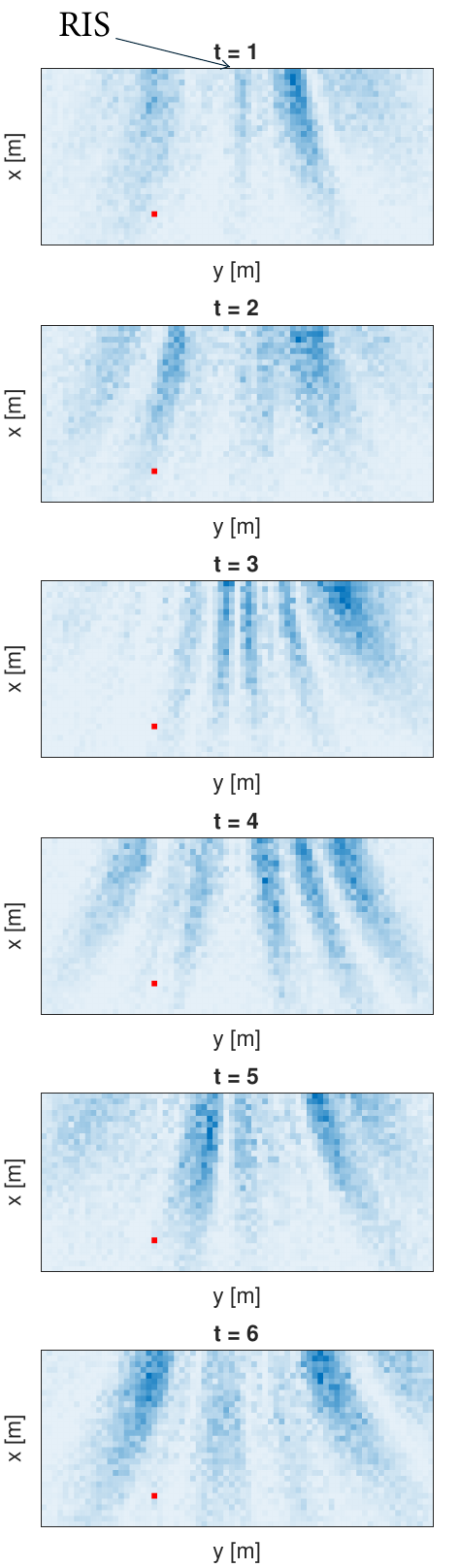}
\caption{Learned RIS (non-active).}%
    \label{loc_ris_interp_6_dnn}
  \end{subfigure}%
  \hfill
\caption{RSS distribution from time frame $1$ to $6$, $T=6$, SNR = $25$dB.}
    \label{loc_ris_interpretation}
\end{figure}

\subsection{Interpretation}

We visually interpret the RIS design obtained from the proposed LSTM based
network. Here, we test the neural network for one single user and use the RSS
distribution (or radio map) as a means to illustrate the beamforming pattern
produced by the adaptively designed RIS reflection coefficients.  To do so, at each time
frame, we record the designed RIS configuration and plot the RSS obtained at
every $1m\times1m$ block in the area of interest across the $x-y$ plane 
as shown in Fig.~\ref{loc_ris_interpretation}. 
The red dot denotes the true UE position.
From Fig.~\ref{loc_ris_interpretation}(\subref{loc_ris_interp_6_rnn}), 
it is clear that the RIS is probing broader beams at the first several time frames to search for the user, then gradually focusing the beam towards the UE 
as $t$ increases. This implies that the proposed neural network is indeed designing meaningful RIS configurations based on the measurements.

\section{Conclusion}
This paper shows that active sensing can significantly improve the performance of
an uplink RIS-assisted localization task in which the BS adaptively designs the 
sequence of RIS reflection coefficients based on the received pilots from the 
user so far to enhance the localization accuracy.  
By employing an LSTM based neural network, the proposed solution
can effectively utilize historical measurements to design RIS configurations
for subsequent measurements in a codeboook-free fashion for the purpose of
minimizing localization error.  Numerical results indicate a lower localization 
error as compared to existing active and nonactive benchmarks. The proposed 
solution demonstrates interpretable results and is generalizable to early stopping in the sequence of sensing stages.


%



\ifCLASSOPTIONcaptionsoff
  \newpage
\fi



%

\bibliography{references}

\bibliographystyle{IEEEtran}

%




\end{document}